\documentclass[aps,prl,twocolumn,superscriptaddress,showpacs]{revtex4}

\usepackage{graphicx}

\begin{document}

\title{Hysteretic magnetotransport in p-type AlGaAs heterostructures with In/Zn/Au ohmic contacts}

\author{B. Grbi\'{c}}
\author{R. Leturcq}
\author{T. Ihn}
\author{K. Ensslin}
\affiliation{Solid State Physics Laboratory, ETH Zurich, 8093 Zurich, Switzerland}
\author{G. Blatter}
\affiliation{Institute of Theoretical Physics, ETH Zurich, 8093 Zurich, Switzerland}
\author{D. Reuter}
\author{A. D. Wieck}
\affiliation{Angewandte Festk\"orperphysik, Ruhr-Universit\"at Bochum, 44780 Bochum, Germany}

\date{\today}

\begin{abstract}
The two-terminal magneto-conductance of a hole gas in C-doped AlGaAs/GaAs heterostructures with ohmic contacts consisting of alloyed In/Zn/Au displays a pronounced hysteresis of the conductance around zero magnetic field. The hysteresis disappears above magnetic fields of around 0.5 T and temperatures above 300 mK. For magnetic fields below 10 mT we observe a pronounced dip in the magneto-conductance. We tentatively discuss these experimental observations in the light of superconductivity of the ohmic contacts.
\end{abstract}


\maketitle

Transport measurements in semiconductors require ohmic contacts between the external metallic leads and the carriers in the semiconductor.  At the interface between a semiconductor and a metal usually a Schottky contact forms giving rise to non-linear current-voltage ($I$-$V$) characteristics. In order to obtain linear $I$-$V$ traces this Schottky barrier has to be overcome, which is usually achieved by highly doping the semiconductor in the contact region. For n-type AlGaAs heterostructures this is routinely achieved by using Au/Ge/Ni alloys. For p-type AlGaAs heterostructures one usually uses layers of AuZn, AuBe or InZn \cite{Williams90}. Especially at liquid He temperatures and below, the quality of the contacts to hole systems is inferior to the ohmic contacts on electron gases. As long as the $I$-$V$ characteristics are linear, high contact resistances can be experimentally overcome by using four-terminal measurements. However, for high impedance devices such as quantum dots one has to perform voltage-biased measurements and the contact quality becomes crucial.

Here we report on a peculiar hysteresis effect observed in two-terminal magnetoconductance measurements on a two-dimensional hole gas contacted by In/Zn/Au alloys. For voltage biased devices the current signal shows a pronounced dip-like feature around zero magnetic field and hysteretic behavior up to magnetic fields where Shubnikov-de Haas (SdH) oscillations become relevant. The feature disappears for high magnetic fields and high temperatures. We tentatively ascribe these observations to the contacts becoming superconducting. This way we extract an estimate for $T_c$, the superconducting transition temperature as well as for critical magnetic fields of the In/Zn/Au alloy which are in tune with values reported in the literature.


Our experiments have been performed on AlGaAs heterostructures doped with carbon (C) acting as an acceptor on (100) substrates \cite{Reuter99,Wieck00}. The integer as well as the fractional quantum Hall effects have been observed in such samples \cite{Grbic04,Grbic05} testifying to the electronic quality of these structures. The material quality has strongly improved over the last couple of years \cite{Gerl05,Manfra05} with mobilities now exceeding $10^6$\,cm$^2$/Vs 
\cite{Zhu07}. Also single electron transistors \cite{Grbic05a} as well as high-quality Aharonov-Bohm rings \cite{Grbic07} have been realized. The effects, which we describe in the following, have been observed in two-terminal measurements on four different samples. We present data from one sample. The wafer consists (from surface to bottom) from 5 nm GaAs, 15 nm Al$_{0.35}$Ga$_{0.65}$As with C doping, 25 nm Al$_{0.35}$Ga$_{0.65}$As, 650 nm GaAs grown on a semi-insulating substrate. At 4.2 K the carrier density is $n = 3.8 \times 10^{11}$\,cm$^{-2}$ and the mobility is $\mu = 120'000$\,cm$^2$/Vs. The sample is patterned into a square mesa ($20\times 20$\,$\mu$m$^2$) geometry by standard photo lithography and etching procedures. Before each evaporation of ohmic contact metallization, the sample surface is cleaned for 2 minutes with oxygen plasma ashing (power 200 W) and the sample is then immediately placed in an evaporation
machine. 

Our experience shows that making ohmic contacts to two-dimensional hole gases (2DHGs) deeper than 100 nm below the surface is relatively straightforward and can be achieved by using only Zn/Au metallization. In the case of shallow 2DHG heterostructures (45 nm below the sample surface) using only Au and Zn did not give satisfactory results.
Therefore we also added In in the contact metallization. We emphasize that the effects reported in this paper were observed only for In/Zn/Au and not for Zn/Au contacts. This
suggests that the presence of In is responsible for the observed effects. We
also found that for a thick initial layer of gold (40 nm) usually used as a sticking layer
the contacts did not work and we had to reduce this thickness to 2
nm. Therefore we use the following contact metallization for
shallow 2DHG samples: 2nm Au, 40nm Zn, 40nm In, 200nm Au annealed
at 480-500$^o$C in 95$\%$ N$_2$ and 5$\%$ H$_2$ atmosphere for 2
minutes. The two-terminal resistances of these contacts are 300-500
k$\Omega$ at room temperature and they drop very fast to values around 10-20 k$\Omega$ upon cooling
the sample to 4.2K. It is
important to point out that the room temperature value of the
two-terminal contact resistance cannot give a clear answer whether the
contacts are good at low temperature. For some samples with
similar values at room temperature, the
contact resistance increased upon cooling to 4.2K. Therefore it is
crucial to check if the contacts are good at 4.2K, before
proceeding with the fabrication of nanostructures.  It is important to note that Indium becomes
superconducting at $T_{c,{\rm In}}=3.4$ K, and zinc at $T_{c,{\rm Zn}}=0.85$
K. Therefore, for experiments in a dilution fridge at
temperatures in the mK range the possible transition of the
Au/Zn/In contact metallization into a superconducting state might
be relevant for transport measurements.

\begin{figure}
\includegraphics[width=3.4in]{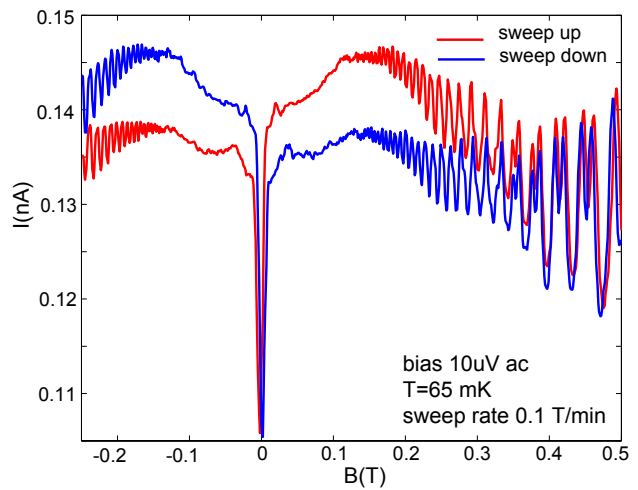}
\caption {Current through a pair of ohmic contacts at T=65mK as a function of magnetic
    field at fixed ac bias of $10 \mu$V and $B$-sweep rate 0.1T/min. The red trace correspond to $B$-field up-sweeps and the blue trace to a down-sweep. As the $B$-field approaches a value around 0.5 T, the hysteresis disappears.\label{fig1}}
\end{figure}
\begin{figure}
\includegraphics[width=3.4in]{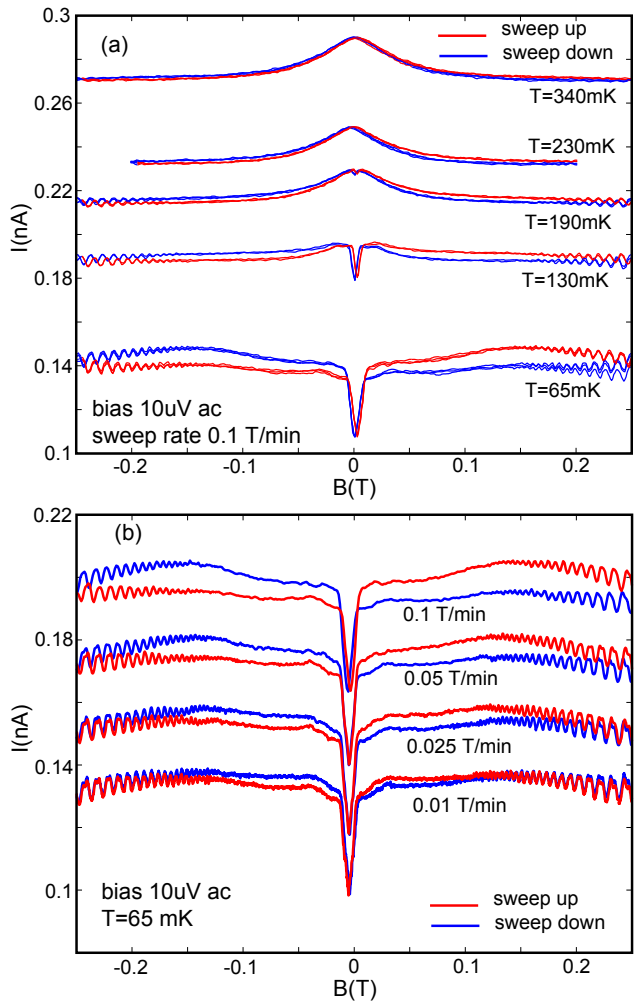}
\caption {(a) Hysteretic magnetoconductance for different temperatures. Red traces correspond to up-sweeps and blue traces to down-sweeps   of the magnetic field. The sweep rate is 0.1T$/$min, and an ac bias of $10 \mu$V is applied across the contacts. The measurements are performed at different temperatures indicated on the right side. Note that the vertical offsets between the traces for different temperatures are not artificially introduced, but originate from the change of contact resistance with temperature.  At each temperature six $B$-field sweeps are performed (three up and three down). The sweeps corresponding to the same
sweep direction lie perfectly on top of each other demonstrating the reproducibility of the measurement. (b) Magnetoresistance for different sweep rates. The traces are vertically offset by 0.02 nA for clarity. \label{fig2}}
\end{figure}

Figure~\ref{fig1} shows a typical two-terminal voltage biased transport measurement in the linear regime on a sample with Au/Zn/In contacts. Figure~\ref{fig2} displays the dependence of the hysteresis on temperature and sweep rate. The observed effects are described in the following:
\begin{enumerate}
\item A hysteresis is observed for up and down sweeps of the magnetic field. A pronounced minimum occurs in the current around $B=0$ with a width of about 10 mT. The conductivity (current) next to the minimum is low before the minimum is reached and high after the minimum has been passed.
\item The hysteresis persists into the regime of Shubnikov-de Haas oscillations and peters out at magnetic fields around 0.5 T.
\item The current dip around $B=0$ as well as the hysteresis weaken with increasing temperature and disappear around 230 mK.
\item The difference between up and down sweeps becomes smaller as the sweep rate is reduced and almost disappears for a sweep-rate of 0.01T/min. The depth of the minimum around $B=0$ does not depend on the sweep rate.
\end{enumerate}

Samples without In in their contacts did not display any of these features. These effects are present only in two-terminal measurements, and they are absent in four-terminal measurements. This indicates that these features originate from the contacts and not from the 2DHG. In particular the zero field
behavior is completely opposite for these two cases. While the two-terminal conductance displays a dip, the four-terminal resistance
displays a minimum due to weak antilocalization \cite{Boris07}, which corresponds to a maximum in the four-terminal conductance.

\begin{figure}
\includegraphics[width=3.4in]{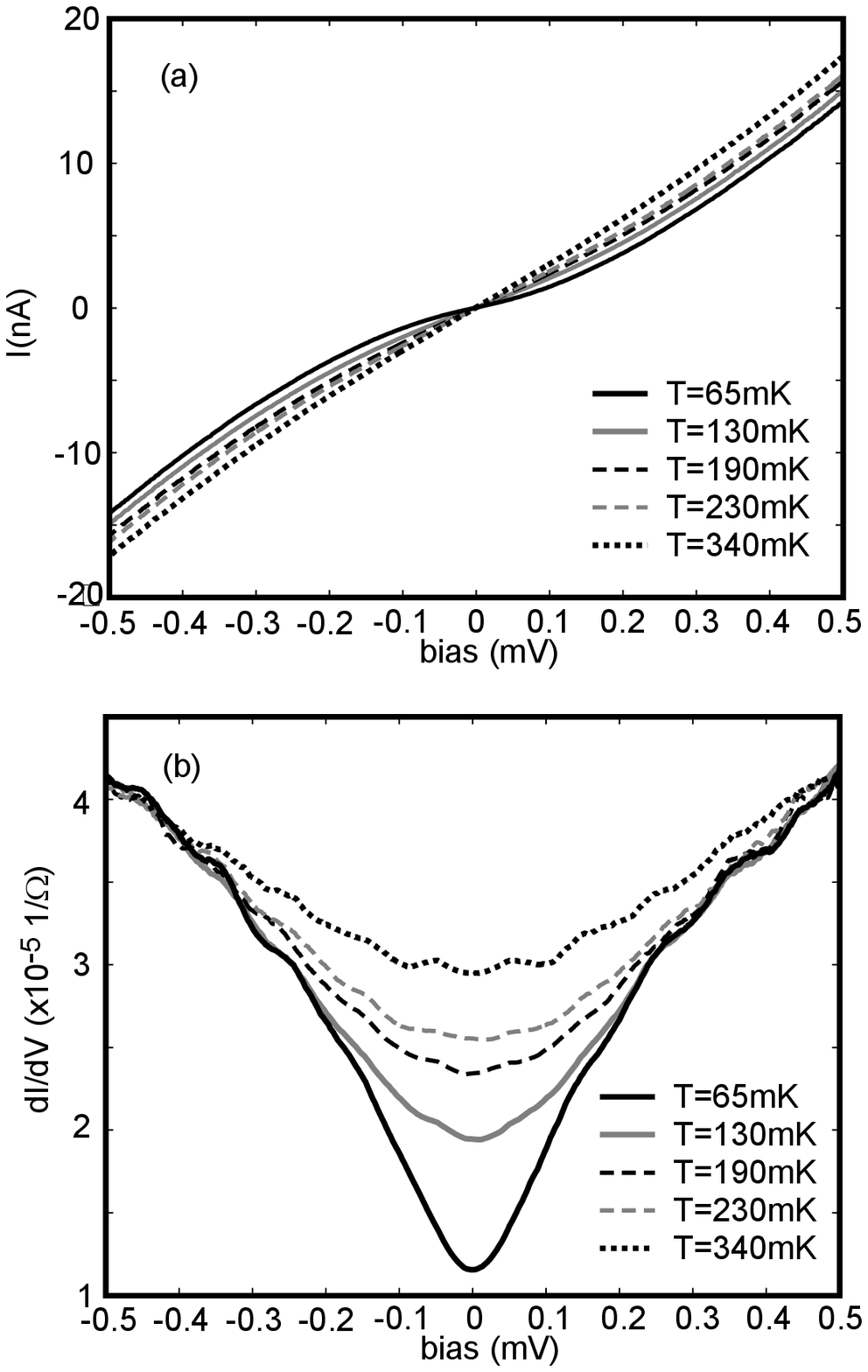}
\caption {(a) Temperature dependence of the $I$-$V$ characteristics of In/Zn/Au ohmic contacts at B=0. (b) Temperature dependence of the differential conductance $dI/dV$, obtained by numerical differentiation of the $I$-$V$ traces shown in (a).\label{fig3}}
\end{figure}

Finally, we have measured the $I$-$V$ characteristics of the ohmic
contacts at different temperatures [Fig. ~\ref{fig3} (a)]. At $T=65$\,mK a
nonlinearity around zero bias is observed, which becomes weaker as
the temperature increases and almost completely disappears above
340 mK. In order to visualize these data in a clearer way
we plot in Fig.~\ref{fig3} (b) the differential conductance $dI/dV$,
obtained by numerical differentiation of the measured $I$-$V$ curves,
as a function of bias. It can be seen that at $T=65$\,mK a very
expressed dip in the differential conduction develops around zero bias with a minimum
around $1\times10^{-5}$ $1/\Omega$. This dip weakens as
the temperature increases to 340 mK. If we analyze this change
quantitatively we see that at $T=340$\,mK, the low-bias differential
resistance is around 30 k$\Omega$ and it increases to around 100
k$\Omega$ as the temperature is reduced to 65 mK. Such a strong
increase of the low-bias two terminal contact resistance as the
temperature is lowered below 100 mK can be very harmful for
low-noise transport measurements, which require low
contact resistances and linear $I$-$V$ charateristics of the contacts.

We like to note that the temperature dependence of the conductance is pronounced for all magnetic fields investigated. The curves in Fig.~\ref{fig2} (a) are not vertically offset. Rather the background conductance changes from about 1/70 k$\Omega^{-1}$ to 1/37 k$\Omega^{-1}$ when the temperature is increased from 65 mK to 340 mK. In the same temperature range the resistance around $B=0$ changes from 90 k$\Omega$ to 35 k$\Omega$ which is consistent with the data of Fig.~\ref{fig3} (b). We conclude that there are two different contributions to the temperature dependence of the resistance, one which is present for the entire magnetic field regime investigated and another one which is particularly pronounced around $B=0$.

These experimental features are linked to the presence of In in the contact
material. Motivated by the temperature and magnetic field dependence of the
observed effects we discuss in the following possible relations
to type II superconductivity in the In/Zn/Au contact pads.

Proximity effects extending between semiconductor
contacts \cite{Nguyen90} have been investigated in InAs-Nb systems where great
care was taken to optimize the interface between the superconductor and the
semiconductor. Indium ohmic contacts were deposited on an n-type AlGaAs
heterostructure at a distance of 1 $\mu$m and the flow of a supercurrent was
demonstrated \cite{Marsh94} and explained in the framework of phase-coherent
Andreev reflections \cite{Marmorkos93}. Once the mobility of the electron gas
was reduced by electron-beam irradiation, a zero-bias dip in the differential
conductance was observed, which was strongly reduced as the magnetic field was
increased above 40 mT. Superconducting Sn/Ti contacts were fabricated close to
a quantum point contact in a high-mobility two-dimensional electron gas in a
AlGaAs heterostructure \cite{Lenssen94}. The authors reported an enhanced weak
localization signal around zero gate voltage. The experimental signatures of
this dip in the magnetoconductance have a similar temperature dependence as
the effects which we report about in this paper. 
The observation of
hysteresis in the conductance upon sweeping the magnetic field was not reported in
these publications \cite{Nguyen90,Marsh94,Lenssen94}.

Since the contacts are annealed into the GaAs host material, we do not expect a
sharp and well-defined interface between the superconductor and the hole gas
in the semiconductor.  Also there is a comparatively large distance (about 250
$\mu$m) between the contacts, much larger than the mean free path, the
inelastic scattering length of the holes in the carrier gas and the pair
coherence length in the superconductor. The experiments were done on a square
mesa and the observed features did not depend on the contact pair used for the
two-terminal experiment. Similar observations were also obtained on another
sample with Hall-bar geometry.

Several experimentally observed features such as two typical magnetic field
scales and a critical temperature indicate a possible relation to the contacts
becoming superconducting. In the following, we discuss possible scenarios and
their limitations to explain the experimental observations.  It is conceivable
that a potential barrier exists for Cooper pairs in the contact region, with a
low transmission to the hole Fermi gas.  One may then attempt to interpret the
above findings by assuming a superconductor-insulator-metal contact with
quenched Andreev scattering:  Charge transmission below the superconducting
gap is suppressed, leading to a small contact conductance at zero magnetic
field. In the presence of a magnetic field, vortices enter the superconductor,
driving the material normal within the core regions of the flux lines.  As a
result, the contact conductance is expected to increase rapidly as vortices
flood the superconducting contacts in agreement with the experimental
observation. A particular feature is the recovery of the zero-field dip in the
current flow upon field decrease, signalling that vortices leave the contact
region as the external magnetic field approaches zero.  This feature is
particular to surface pinning through Bean-Livingston barriers \cite{Bean64}
or due to geometrical barriers \cite{ZeldovLGKMKVS_94} --- in both cases
vortices do leave the superconductor smoothly as the field vanishes.  Beyond
this sharp minimum, which is nearly reversible, the magnetoconductivity
becomes strongly hysteretic, as displayed in Figs.\ \ref{fig1} and \ref{fig2}. It is here
that the above scenario fails: for a given magnetic field one expects that the
number of vortices and therefore the current is smaller/larger for
increasing/decreasing fields, a simple consequence of the magnetic behavior of
a type II superconductor with surface pinning, cf.\ Ref.\ \cite{Burlachkov93}.
This is opposite to the experimental observation.

If on the other hand the contacts between superconductor and electron gas are
assumed to be of good quality without an insulator in between, then the high
current state would be related to the situation with few vortices. Such a
scenario would better explain the hysteretic behavior at finite magnetic
fields, but fails to account for the current minimum at $B=0$.

The sweep-rate dependence of the hysteresis indicates the presence of vortex
creep.  Relaxation to the equilibrium magnetization should then be asymmetric
\cite{Burlachkov93}, with the high current branch (many vortices) decaying
slower than the low current one (fewer vortices).  From Fig.~\ref{fig2} (b) we
learn that the curves swept from high fields towards $B=0$ (red curves for
$B<0$ and blue curves for $B>0$) are equidistantly spaced in the vertical
direction, i.e., they are independent of sweep velocity. On the contrary the
sweeps from low fields to high fields (blue curves for $B<0$ and red curves
for $B>0$) depend strongly on the sweep rate and approach their lower-lying counterpart as the sweep rate is
reduced.  This experimental observation is in contrast to the theoretical
scenario described before.

Since all the observed effects disappear above about 300 mK, this temperature
can be interpreted as the critical temperature of the superconductor.  The
values for the critical fields as well as for the critical temperature $T_c$
are not unreasonable considering the numbers known in the literature
\cite{numbers,Martinoli71}. While the likelihood that the observed effects are
related to the superconductivity of the contacts is large, we conclude,
however, that the microscopic understanding of the contact region and the
transport mechanism is lacking.

The flat region of the $I$-$V$ curve in Fig. ~\ref{fig3} (a) probably arises
from two different effects. One is a simple heating effect which we have also
observed for other p-type nanostructures. This effect appears to be very
significant in the case of holes.  Heating effects may explain the bias
dependence of the conductance over the entire magnetic field range
investigated. Another weaker bias dependence of the conductivity
(Fig.~\ref{fig2} (a)) could be related to the superconductivity of the
contacts. Because the two effects occur simultaneously it is not
possible to estimate a gap energy from the non-linear $I$-$V$ characteristics
at low temperatures.

We refrain from a more detailed and elaborate interpretation of the data in
view of many experimental uncertainties. The superconductivity of the contacts
has been discussed as a possible origin for the observed features. For
experimentalists working with nanostructures in semiconductors it is crucial
to recall the importance of ohmic contacts in general and in particular for
two-terminal transport measurements.

%
%

\begin{acknowledgments}
Financial support from the Swiss Science Foundation (Schweizerischer Nationalfonds) is gratefully acknowledged.
\end{acknowledgments}



\begin{thebibliography}{18}
\bibitem{numbers} AuIn$_2$ is a type I superconductor with $T_c=0.22$ K and
$B_c=1.7$ mT. AuIn$_3$ is a type II superconductor with $T_c=50~\mu$K.
Au$_x$Zn$_{(1-x)}$ is a type II superconductor with $T_c=0.56-1.21$ K.
\end{thebibliography}
\end{document}